# Enhanced room temperature ferromagnetism in nanostructured MoS$_2$ flakes by hydrogen post-treatment: Combined experimental and first-principles-based studies

*Sharmistha Dey[a], Ankita Phutela[b], Saswata Bhattacharya[b], Fouran Singh[c], Pankaj Srivastava[a], and Santanu Ghosh[a*]*

[a] Nanostech Laboratory, Department of Physics, Indian Institute of Technology Delhi, New Delhi 110016, India

[b] Department of Physics, Indian Institute of Technology Delhi, New Delhi 110016, India

[c] Materials Science Group, Inter-University Accelerator Centre, New Delhi 110067, India

**Abstract:**

We meticulously study the individual effects of hydrogen irradiation and annealing on the electronic structure and magnetic properties of nanostructured MoS$_2$ thin films grown through Chemical Vapor Deposition (CVD). The role of edge-terminated structure and point defects to induce room-temperature ferromagnetism (RTFM) is thoroughly investigated. The nanostructured pristine MoS$_2$ thin films show the formation of a pure 2-H MoS$_2$ phase, confirmed by X-ray diffraction (XRD) and Raman spectroscopy. Pristine MoS$_2$ thin films are independently annealed in a reducing hydrogen environment and irradiated with low-energy hydrogen ions to study the significance of point defects like sulfur vacancies. RTFM with saturation magnetization value (M$_s$: 1.66 emu/g) has been observed in the pristine film. Magnetization increases after irradiation and annealing processes. However, hydrogen annealing at a temperature of 200°C exhibits a maximum M$_s$ value of 2.7 emu/g at room temperature. The increase in ferromagnetism is attributed to an increment in sulfur vacancies, hydrogen adsorption with sulfur, and modification in edges, which is confirmed by the analysis of Electron probe micro-analyzer (EPMA), X-ray photoelectron spectroscopy (XPS), and Field emission scanning electron microscopy (FESEM) measurements. The Density Functional Theory (DFT) calculations have demonstrated that the edge-oriented structure of MoS$_2$ exhibits a magnetization value of 3.2 μB. Additionally, introducing an S-vacancy and H-adsorption in a parallel position to the sulfur further enhances the magnetization value to 3.85 μB and 3.43 μB respectively. These findings align broadly with our experimental results.

**Corresponding author:** santanu1@physics.iitd.ac.in

# 1. Introduction:

Ferromagnetic semiconductors are crucial for spintronics, as both electronic charge and spin are required simultaneously for such applications[1,2]. Semiconductors are typically diamagnetic at ambient temperature. To make the semiconductor ferromagnetic, transition metal doping-induced magnetism is explored[3–6]. Even though there is a substantial amount of magnetization induced, it is never clear if the magnetism originates intrinsically or due to the agglomeration of transition metals. To eliminate this, defect-induced room temperature ferromagnetism has been started to be elaborately investigated[7–11]. Disruption of lattice symmetry and corresponding changes in the electrical, optical, and magnetic properties can all be caused by defects that are not present in ideal materials[12]. Defect-induced magnetism is reported in graphene at room temperature, and it provides a new research perspective for the researchers[13–15]. As two-dimensional (2D) transition metal dichalcogenides (TMD) resemble graphene, there is interest in investigating the defect-induced magnetism in nanostructured TMD semiconductors.

Among TMDs, Molybdenum disulfide ($MoS_2$) has garnered much interest due to its unique characteristics, such as effective catalytic[16], photoconductive[17], optoelectronic[18], and lubricant properties[19]. It has a layered structure, a thickness-dependent bandgap (1.9 eV direct band gap for monolayer and 1.2 eV indirect bandgap for bulk), inter-layer weak Van der Waals interaction, and an intra-layer strong covalent bond[20]. Bulk $MoS_2$ is a diamagnetic semiconductor, but as nanostructured edge-terminated $MoS_2$ thin film has a different coordination geometry concerning the bulk, it shows ferromagnetic behavior at room temperature[21]. To differentiate the magnetic behavior between bulk and nanostructured materials, researchers have recently focused on this field of research. Apart from edge states, there are different types of inherent defects present in $MoS_2$, including S vacancy, Mo vacancy, S substitute Mo vacancy, and Mo substitute S vacancy, etc[22]. If RTFM can be triggered due to edge states and point defects or a combination of both, it will be a potential material for spintronics applications. Few studies have been reported on defect-induced magnetism in the $MoS_2$ monolayer, a few layers, and bulk[7–9,11,23–26]. Isolated vacancies, vacancy clusters, generation and reconstruction of edge states, defects, lattice distortion, etc. have been mentioned as the reasons for the induced RTFM.

To create different types of defects and change structural and other physical properties, low-energy ion-irradiation and annealing in a reducing environment are employed. In low-energy ion-irradiation regimes, ions transfer their energy through elastic collisions, known as nuclear energy loss ($S_n$), which causes various point defects, including vacancies and interstitials, as a result of the collision cascade[27]. As $MoS_2$ has a layered structure and van der Waals interaction between two layers, irradiation with low-energy light mass ions would be effective in generating defects. Heavy mass ion irradiation can cause unfavorable deterioration of their physical properties[28]. The creation of S-vacancies and modification of physical properties occur in the case of annealing in $H_2$ ambient at an optimum temperature [29,30].

The present study originates with the motivation to understand the origin of RTFM and its modification in the nanostructured pristine, and hydrogenated defect-engineered $MoS_2$. CVD-grown nanostructured $MoS_2$ thin films were prepared by optimizing all the growth parameters in a single-zone programmable tubular furnace. Pristine nanostructured $MoS_2$ thin films exhibit ferromagnetism at room temperature. To create defects like S-vacancies and enhance the magnetization, 30 keV hydrogen ion irradiation with different fluences such as $5\times10^{13}$, $5\times10^{14}$, and $5\times10^{15}$ ions/cm$^2$ and hydrogen annealing at different temperatures such as 100ºC, 200ºC, 300ºC, for 1 hour were performed. In both cases, the magnetic moment is enhanced with respect to the pristine sample. Out of all the cases studied here, the highest magnetization is observed for the 200ºC $H_2$-annealed sample. The following physical processes have been discussed to understand the induced magnetism (i) edge-terminated nanostructure, which contains different stoichiometry than bulk; (ii) the creation of defects like S-vacancies, hydrogen adsorption with sulfur, and active sites in the basal plane. A correlation is made with density functional theory (DFT) based calculation. This article presents a comprehensive analysis of the process of tuning magnetism in nanostructured $MoS_2$ by hydrogen irradiation and annealing, including its correlation with electronic structure and DFT-based theoretical calculations. To the best of our knowledge, the present study shows maximum saturation magnetization at room temperature in $MoS_2$ as compared to previously reported values (Table S1).

## 2. Experimental and computational details:

### 2.1. Experimental:

Vertically nanostructured $MoS_2$ thin films were grown on $SiO_2$/Si substrates by optimizing all the chemical vapor deposition (CVD) growth parameters such as precursor quantity, growth temperature, substrate type, the distance between precursors and substrates, and deposition time, etc. respectively, according to a single zone programmable tube furnace. Molybdenum trioxide ($MoO_3$, 15mg) and sulfur (S, 150mg) powder were used as precursors and kept in a 3 cm diameter and 90 cm long quartz tube placed inside the tubular furnace. $MoO_3$ powder was used as a molybdenum source, placed in the middle part (hottest region, 850ºC), and S powder was kept 26 cm away from the $MoO_3$ powder at a lower temperature zone ( 120ºC) of the quartz tube. 50 sccm Argon gas flow was used as a carrier gas and controlled by a mass flow controller. Ar gas was flown for 30 minutes before the deposition to establish an inert environment. A cleaned $SiO_2$/Si substrate was kept face up at a distance of 12 cm away from the MoO3 powder, where the temperature was 550°C and at a height of 1.2 cm from the bottom (shown in Figure 1a). Subsequently, the furnace was heated up to 850°C at a ramp rate of 12°C per minute. The deposition time was 20 minutes after reaching the desired temperature. The deposited film was allowed to cool naturally. A pure-phase vertically nanostructured MoS$_2$ thin film was obtained on a large area. At first, S vaporized in the upstream region and flowed with the Ar gas, reaching the central part where MoO$_3$ started to vaporize. Sulfur reacts with MoO$_3$ and forms MoO$_2$, as the quantity of S is higher than the MoO$_3$ powder, then it reacts with MoO$_2$ and forms MoOS$_2$, and finally excess S reacts with MoOS$_2$ to form MoS$_2$[20,31–33].

The related chemical reactions are:

1. $2MoO_3 + S \rightarrow MoO_2 + SO_2\uparrow$

2. $2MoO_2 + 5S \rightarrow 2 MoOS_2 + SO_2\uparrow$

3. $2 MoOS_2 + S \rightarrow 2 MoS_2 + SO_2\uparrow$

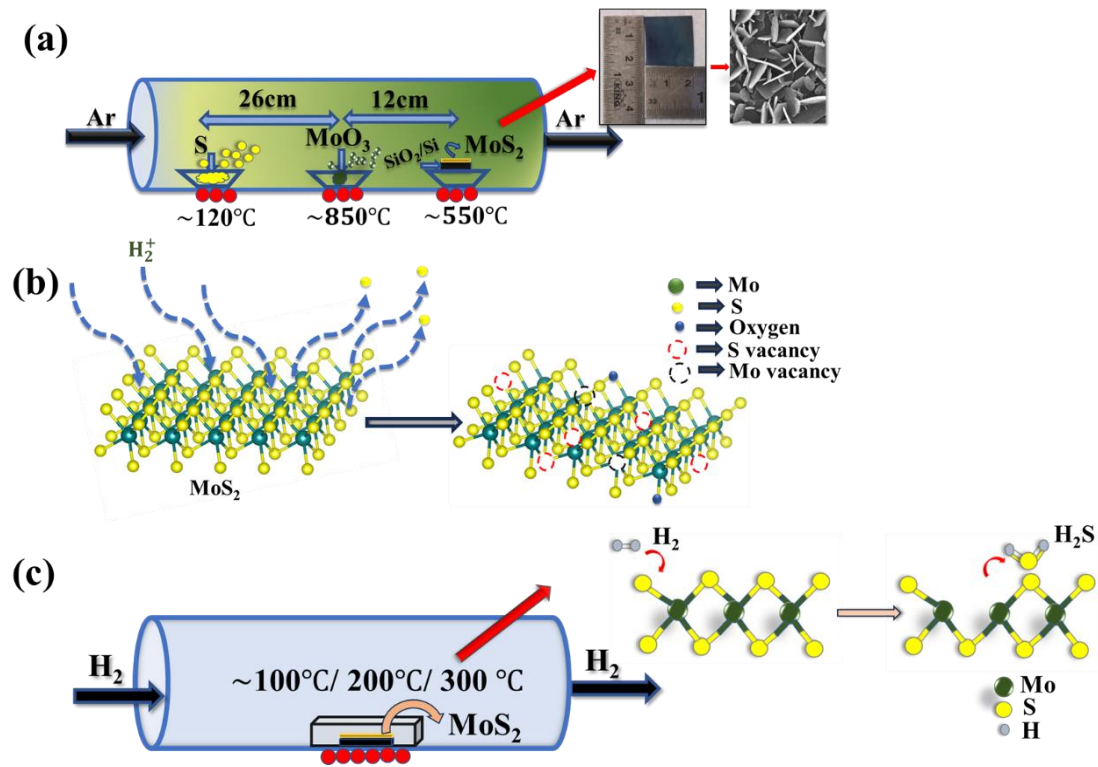

**Figure 1.** (a) Schematic diagram for the CVD setup for the deposition of MoS$_2$ nanostructured thin films. Deposition was obtained on a 2×2 cm$^2$ substrate (see inset). (b) Schematic diagram of H$_2^+$ ion irradiation on MoS$_2$ and creation of S vacancies. (c) Schematic diagram of H$_2$ annealing in a single zone furnace at 100°C, 200°C, and 300°C temperatures for one hour and the creation of S vacancies.

## 2.2. Computational:

The Vienna *ab initio* simulation package (VASP)[35], utilizing projected augmented wave (PAW) potentials[36], was employed for conducting first-principles calculations within the framework of density functional theory (DFT)[37,38]. The exchange-correlation interactions among electrons were treated using the generalized gradient approximation (GGA) with the Perdew-Burke-Ernzerhof (PBE) functional[39]. To ensure structural accuracy, the atomic positions were relaxed until the Hellmann-Feynman forces on each atom were minimized to less than $10^{-5}$ eV/Å. The electronic self-consistency loop convergence criterion was set to $10^{-5}$ eV, and a kinetic energy cutoff of 520 eV was chosen for the plane-wave basis set expansion. The Brillouin zone was sampled using a Γ-centered 1×4×1 k-grid. To prevent electrostatic interactions between periodic images, a vacuum space of 20 Å (in z-direction) and 8.7 Å (in x-direction) was incorporated around all edges of the simulated structures. Additionally, the two-body Tkatchenko-Scheffler van der Waals correction scheme was applied to optimize the structures[40,41].

## 3. Results and discussions:

Figure 2 shows the XRD pattern of pristine, $H_2^+$ irradiated, and $H_2$ annealed $MoS_2$ samples deposited on $SiO_2$/Si substrates at room temperature. The peak at ~14.36° is corresponding to the (002) plane of $MoS_2$[7,42–44]. As no other peak is observed, it indicates that the film has a c-axis orientation. It also confirmed the pure 2-H phase formation of $MoS_2$ nanostructured thin films with trigonal prismatic coordination geometry with space group $P6_3/mmc$[45]. No other phases or impurity phases are observed in the XRD pattern for the pristine and post-treated samples. From Figure 2a, it is evident that after $H_2^+$ irradiation there is a shift of the (002) peak in a lower Bragg angle from 14.36° to 14.19° for the H0 to H3 samples. An enlarged view of the red shift of the peak is shown in Figure 2b. The FWHM also gradually increases from 0.77° to 1.25° after irradiation from the pristine to H3 sample. A similar pattern was observed for the $H_2$ annealed samples (Figure 2c). It is evident that after $H_2$ annealing there is also a shift towards a lower diffraction angle of the (002) peak from 14.36° to 14.13° for the H0 to HC samples. An enlarged view of the red shift of the peak is shown in Figure 2d. The FWHM also increases after annealing from 0.77° to 1.11° from the pristine to HC sample. An increase in FWHM after irradiation and annealing manifests the creation of defects, like vacancies, and deterioration in crystallinity. The observed redshift on irradiation and annealing indicates the development of micro-strain and swelling of the lattice[11]. The average crystallite size (D) is calculated from Scherrer's formula corresponding to the (002) plane of $MoS_2$. The crystallite sizes of H0, H1, H2, H3, HA, HB, and HC samples are found to be 10.3, 9.1, 8.3, 6.3, 8.9, 7.4, and 7 nm respectively. The lowering of crystallite sizes after hydrogen irradiation and annealing suggests deterioration and damage in the lattice[11,46,47].

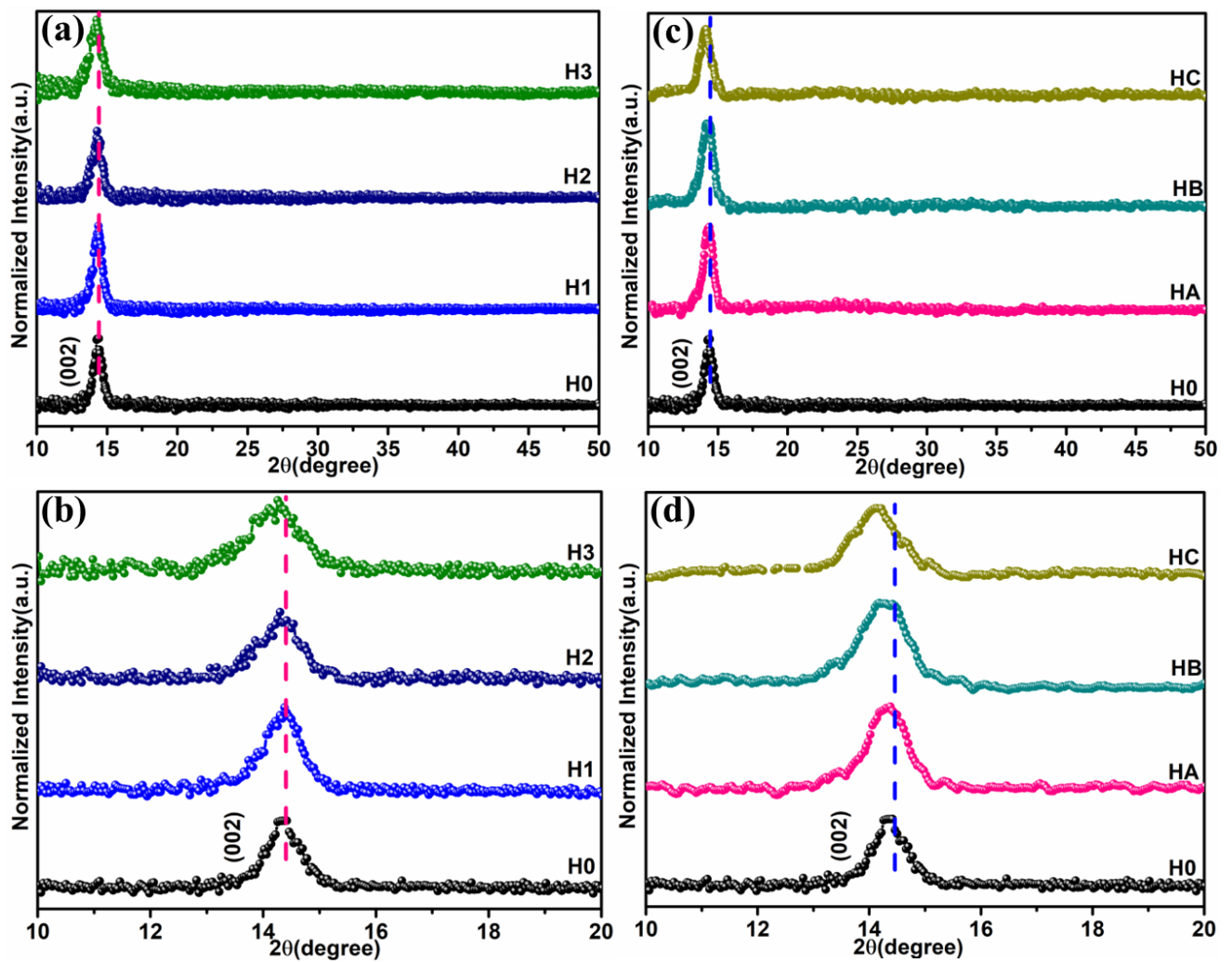

**Figure 2.** XRD pattern of (a) Pristine and $H_2^+$ irradiated samples; (b) Enlarged view and left shift of (002) peak for the irradiated samples; (c) Pristine and $H_2$ annealed $MoS_2$ samples; (d) Enlarged view and left shift of (002) peak for the annealed samples. For both cases, there is a shift towards the lower Bragg angle in the position of the (002) peak and an increase in FWHM with increasing irradiation fluences, confirming lattice degradation and the creation of defects.

Figure 3 exhibits the Raman spectra of pristine, $H_2^+$ irradiated, and $H_2$ annealed samples at room temperature. The peak located at 383.5 cm$^{-1}$ is denoted by $E_{2g}^1$ peak (in-plane vibration of Mo and S atoms), and at 409.3 cm$^{-1}$ is denoted by $A_{1g}$ peak (out-of-plane vibration of S atoms). The peak situated at 452 cm$^{-1}$ is assigned to the second-order longitudinal acoustic (2LA) mode [23,48], originating due to the double resonance Raman process. There is no prominent change in 2LA for both cases, as it is not directly correlated to the defect formation[49,50]. From Figure 3a, it is evident that after $H_2^+$ irradiation, there is a redshift of the $E^1_{2g}$ and $A_{1g}$ peaks from 383.5 to 381.9 cm$^{-1}$, and from 409.3 to 407.9 cm$^{-1}$, respectively. Also, the intensity of these two peaks decreases from the pristine to the H3 sample. The FWHM also increases after irradiation (For the $E^1_{2g}$ peak, it increases from 7.69 to 10.21 cm$^{-1}$, and for the $A_{1g}$ peak, it increases from 7.25 to 8.29 cm$^{-1}$ from the pristine to the H3 sample). A similar pattern was observed for the H$_2$-annealed samples, which is clearly shown in Figure 3b. It is evident that after H$_2$ annealing, there is also a redshift of the $E^1_{2g}$ and $A_{1g}$ peaks from 383.5 to 382.1 cm$^{-1}$, and from 409.3 to 408.2 cm$^{-1}$, respectively. Also, the intensity of these two peaks decreases from the pristine to the HC sample. An enlarged view of the shifting of the peaks for both cases is shown in Figure S1. The FWHM also increases after annealing (for the $E^1_{2g}$ peak, it increases from 7.69 to 10.1 cm$^{-1}$, and for the $A_{1g}$ peak, it increases from 7.25 to 8.4 cm$^{-1}$ from the pristine to HC sample). The decrease in intensity and an increase in FWHM after both post-treatments manifest that there is an induced tensile strain (stretching in bond length) due to the generation of different types of vacancies (like S-vacancies) and defects[7,11,28,29,34]. The $E^1_{2g}$ peak is more sensitive (Figure S1) to the creation of vacancies and other types of defects than the peak $A_{1g}$, as the length of the covalent bond between two nearby Mo atoms decreases due to the presence of S-vacancies, leading to a more pronounced decrease in the in-plane vibration energy compared to the out-of-plane vibration[50,51].

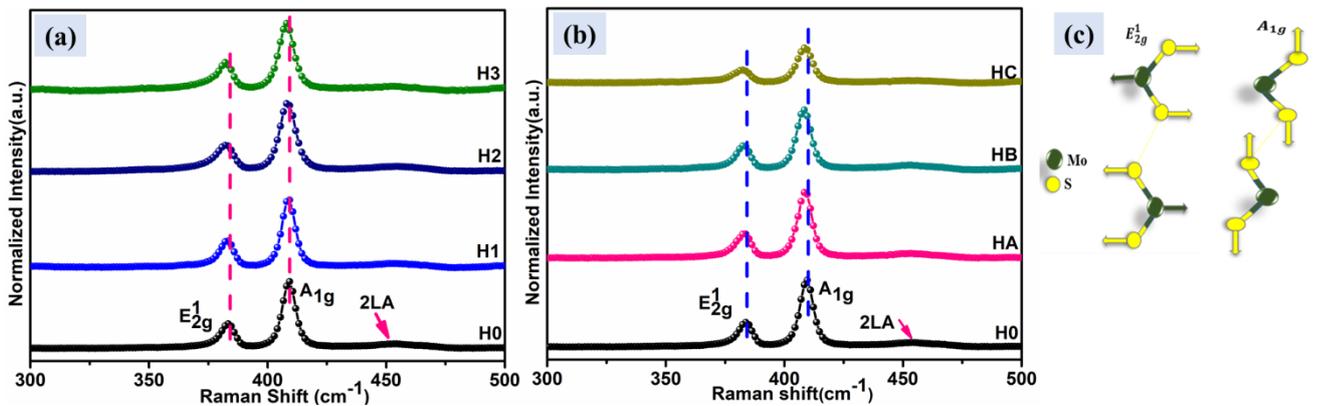

**Figure 3.** Raman spectra of (a) Pristine and H$_2^+$ irradiated samples; (b) Pristine and H$_2$-annealed MoS$_2$ samples; (c) Schematic diagram of in-plane vibration of S and Mo atoms ($E^1_{2g}$ mode) and out-of-plane vibration of S atoms ($A_{1g}$ mode).

The surface morphology of pristine, hydrogen-irradiated, and hydrogen-annealed samples was examined using FESEM (Figure 4). As can be seen, the edge-terminated nanostructured MoS$_2$ film shows changes after hydrogenation post-treatment. Hydrogenated post-treatment has already been implemented to generate nanoribbons of graphene to increase the active sites[29,52]. Figure 4 shows the modification of the nanosheets after post-treatment. The H0 sample contains vertical nanosheets with edge terminations. Some vertical nanosheets have started to fragment into smaller ones, increasing the number of edges from pristine to HA and H1 samples. On further increasing the irradiation fluences and annealing temperature, some vertical nanosheets deteriorate (shown by the yellow square in the Figure), whereas others remain intact. After that, the edge of the nanosheets and the basal plane become rough, indicating increasing the active sites[34]. For H3 and HC samples, nanosheets are damaged and agglomerated, decreasing the number of edges and also vacancies, as there is a chance to incorporate sputtered sulfur into the vacancies. As unsaturated spin at the prismatic edges is the main reason behind the magnetism in nanostructured $MoS_2$ thin films, thick films contain a smaller number of prismatic edges due to the secondary nucleation of these edges[21]. The thickness of the nanosheets for the H0 sample is 25nm (Figure S2), which gets sharpened after post-treatment (for HB and H2 samples), which further enhances the induced magnetization.

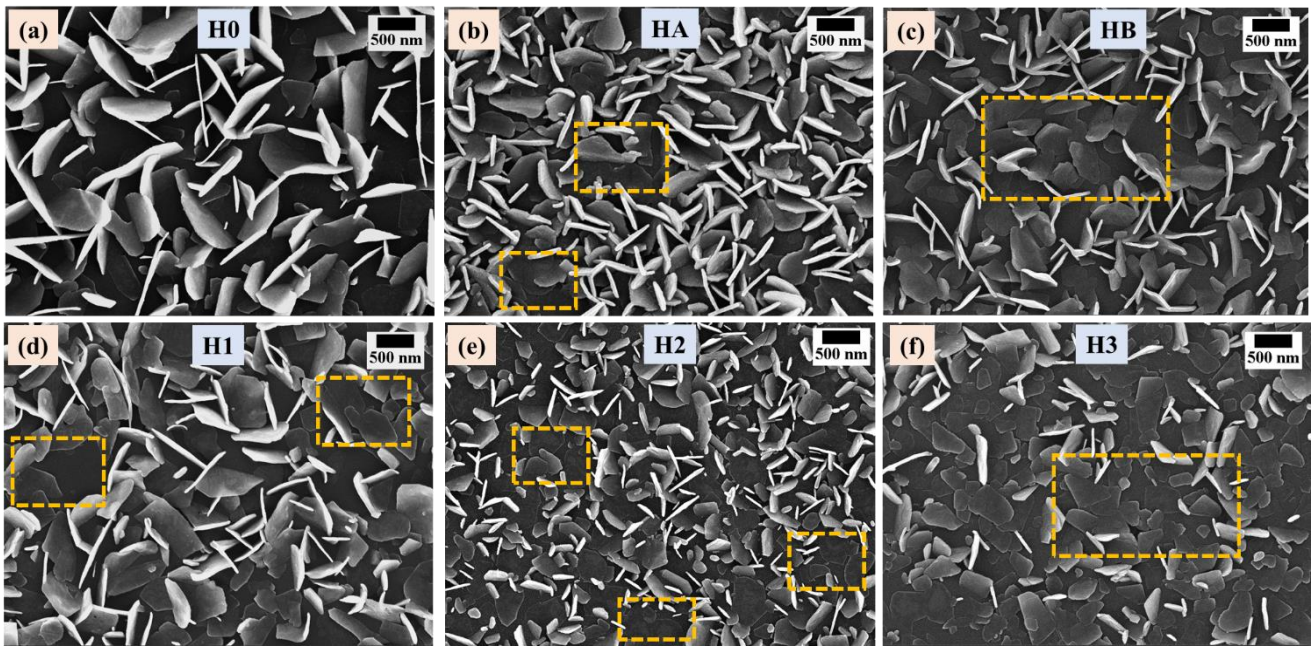

**Figure 4.** FESEM images of (a)-(c) Pristine and $H_2$-annealed samples and (d)-(f) $H_2^+$ irradiated samples. It is clear from the Figures that for both cases there is morphological evolution in the samples.

Electron probe micro analyzer (EPMA) analysis is performed for the mapping of the samples in a selected area to show the uniform distribution of Mo and S over the sample and also to confirm the absence of any magnetic impurities from the qualitative analysis (Intensity versus wavelength plots)[33]. Here, mapping and qualitative analysis are performed for the H0, H2, and HB samples. Mapping and qualitative analysis are shown for the HB sample in Figure 5 on two different points (Figures S3 and S4 for H0 and H2 samples), and it confirms the presence of only Mo and S in all the samples with no magnetic impurity. The intense Silicon (Si) signal is coming from the substrate and has no magnetic contribution. At first, an area is selected, which is shown in Figure 5(a), and mapping is carried out independently for Mo and S elements. For Mo and S detection, a PET (Pentaerythritol) crystal detector is used. Figure 5(b) shows the mapping for S, and Figure 5(c) shows the mapping for Mo. Figures 5(d) and (e) show the qualitative analysis for Mo and S at two different points. It is evident that after ion irradiation and also annealing, the relative S/Mo ratio decreases from 1.82 to 1.23 and 1.15 from H0 to H2, and HB samples, confirming the creation of S vacancies.

Figure 6 shows the magnetization (M) versus applied magnetic field (H) plots for the pristine, hydrogen-irradiated, and hydrogen-annealed samples at room temperature. The M versus H plots for all the samples recorded at 2K are shown in Figure S5. Figure 6a shows that pristine and all hydrogen-irradiated samples show a clear hysteresis loop at 300K, but the saturation magnetization value decreases for the H3 sample. The enlarged view of the central part is shown in the inset of Figure 6. From Figure 6b, it is evident that samples H0, HA, and HB show a clear hysteresis loop, but at a higher temperature (300°C), the annealed sample (HC) shows diamagnetic behavior. The saturation magnetization values for H0, H1, H2, H3, HA, and HB samples are 1.66, 1.87, 2.39, 1.06, 2.18, and 2.7 emu/g at room temperature, respectively. The values of saturation magnetization ($M_s$), coercive field ($H_c$), and remnant magnetization ($M_r$) at 300K are tabulated in Table S2 and for 2K in Table S3. Comparing pristine and HB samples, an enhancement of about 63% in the Ms value is observed. Robust Ferromagnetism is observed from 2K up to ambient temperature, and an increase in coercivity at low temperatures also indicates the ferromagnetic behavior of the samples. Magnetization versus temperature measurement is performed in a standard oven set up from 300 to 900K at a 1000 Oe field for the pristine sample, which is shown in Figure S5.c. It is evident from the M versus T curve that the transition temperature is 830K, i.e., nanostructured $MoS_2$ thin film is ferromagnetic at room temperature.

Transition metal-doped $MoS_2$ exhibits ferromagnetism at ambient temperature[3,5,6,53,54], however, the specific origin of the induced magnetism, whether it be the agglomeration of the transition metal or the synergistic effect of $MoS_2$ and dopants, is continually a matter of debate. Therefore, the magnetism induced by defects was investigated. Possible causes of the induced magnetism in $MoS_2$ include reconstruction of the lattices, isolated vacancies or vacancy clusters, and edge states[9]. Additional reports have demonstrated that the induction of zig-zag edges and defects can produce room-temperature ferromagnetism in nanocrystalline $MoS_2$ films and nanoribbons[8,55]. A correlation between strain tuning and defects-induced magnetism is rarely reported in some theoretical[56,57] and experimental investigations[11,58]. The bound magnetic polaron (BMP) induced by the spin-interaction of confined electrons of S-vacancies and Mo 4d ions was also used to explain ferromagnetism[7]. The magnetic moments obtained in various research studies concerning $MoS_2$ by proton irradiation are compiled in Table S1. There are a few reports on proton irradiation on $MoS_2$, which suggest that low-energy proton and hydrogen irradiation create vacancies (mainly S-vacancies), point defects, and surface reconstruction. Also, the relative percentage of Mo 6+ decreases as compared to the 4+ states with increasing irradiation fluences, which increases saturation magnetization (discussed in the later section)[7,9,11,59].

In the present work, nanostructured, pristine $MoS_2$ thin films show RTFM with a moment of 1.66 emu/g. The unsaturated Mo atoms at the edge states and S vacancies are the main reasons behind the induced magnetism. The edge-terminated atoms have different stoichiometry than the bulk atoms, and Mo atoms will be unsaturated at the edges with octahedral coordination, which will result in spin polarization and induced ferromagnetism. The higher the number of unsaturated prismatic edges, the greater the enhancement in ferromagnetism. There is the formation of S vacancies, as evident from EPMA and XPS data (discussed in the later section) after hydrogen irradiation and annealing. The highest magnetization observed for the HB sample is 2.7 emu/g. However, as $MoS_2$ is a layered structured material, it tends to absorb oxygen at the surface, so the pristine sample contains some Mo 6+ states. After irradiation and annealing in a reductive environment, Mo 6+ states converted into 5+ states and also created adsorption of hydrogen with sulfur, S-vacancies, and metallic Mo, causing an increasing magnetization value. To confirm this experimentally, one pristine sample was annealed at 100°C for 2 hours in the presence of 50 sccm oxygen, and magnetic measurements were performed. It has been noticed that the magnetization value decreases from 1.66 emu/g to 0.29 emu/g for the pristine to $O_2$-annealed sample (Figure S5.d). A detailed explanation is given in the XPS section. The reason behind the enhancement in the magnetism is qualitatively understood based on the following two processes: (i) The creation of active sites, adsorption of hydrogen with sulfur, and S-vacancies in the basal plane due to irradiation and annealing. The spin polarization of the unpaired electrons corresponding to Mo around the S-vacancy generates induced magnetic moments that exhibit ferromagnetism through cooperative interaction. (ii) the edge-terminated structure, which also gets roughened and creates active sites.

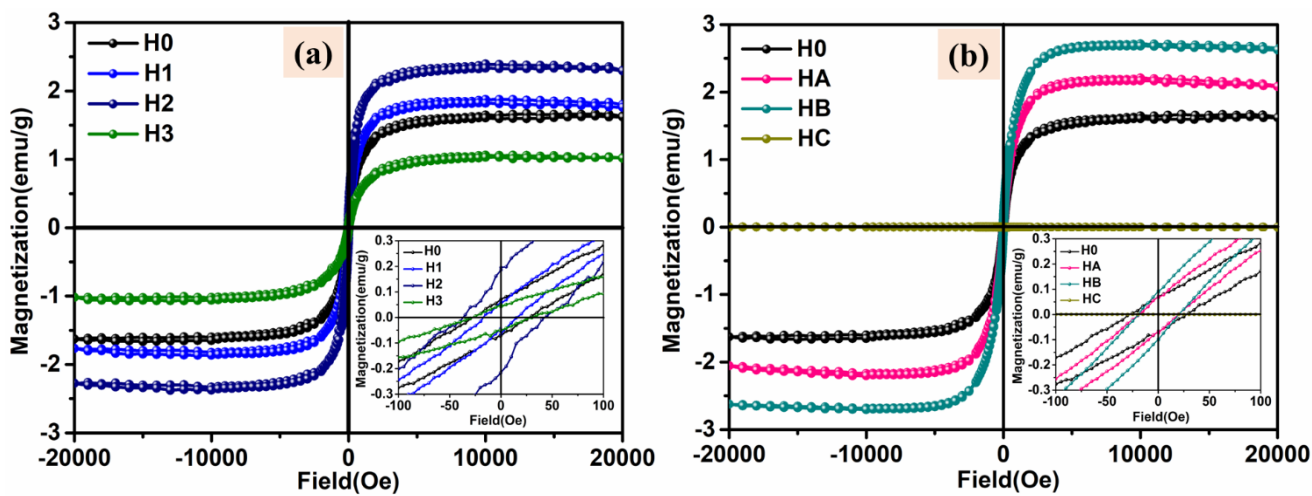

**Figure 6.** (a) M versus H plots for pristine and $H_2^+$ irradiated samples at room temperature; (b) M versus H plots for pristine and $H_2$ annealed samples at room temperature. All samples, except HC, show a hysteresis loop at room temperature. HC sample exhibits a diamagnetic nature.

External influences have a distinctive effect on the electronic structure of nanostructured two-dimensional materials[60,61]. To elucidate the rationale for the enhancement in magnetization and elemental composition at the surface, the electronic structure of hydrogen-irradiated and annealed samples as obtained from the XPS measurements has been elaborately studied. Figure 7 presents a stacked representation of the Mo 3d and S 2p core-level XPS spectra of $MoS_2$ pristine and $H_2^+$ irradiated samples. It is evident that after irradiation, all the peaks exhibit a shift towards lower binding energy (the binding energy shifted from 229.6 eV to 229.1 eV from H1 to H3 samples for Mo 3d spectra). Additionally, peak broadening in the Mo 3d and S 2p core-level spectra is observed, which is indicative of the formation of defects, vacancies, surface amorphization, and disorder[62–64]. A similar pattern is observed for the $H_2$-annealed samples (Figure S6). The samples contain no discernible transition metal impurities, which is confirmed by the XPS survey spectrum of the pristine sample (Figure S7). The peak shift is clearly visible from H0 to H2 samples (from H0 to HB samples in Figure S6). There is no significant shift in the H3 sample as compared to the H2 sample, as at the highest fluence there is oxygen and excess sputtered sulfur incorporation in the vacancy sites (elaborately studied in the next section).

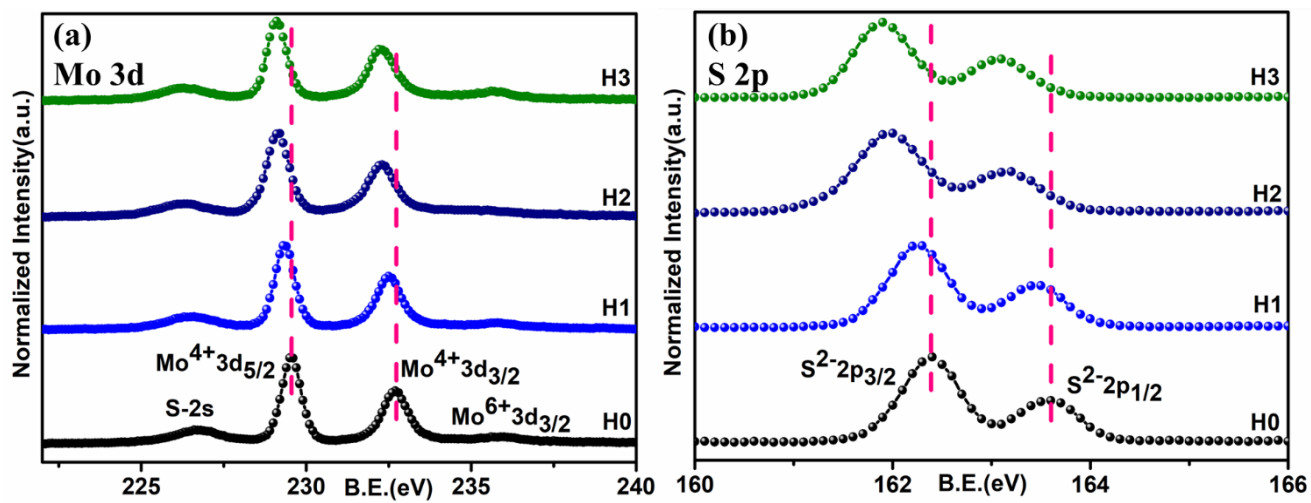

**Figure 7.** (a) Mo 3d, and (b) S 2p core level XPS spectra of pristine and $H_2^+$ irradiated samples. A shift towards lower binding energy in both Mo 3d and S 2p spectra confirms the formation of S-vacancies.

In Figures 8 and 9, the Mo 3d core-level XPS spectra are deconvoluted into nine peaks, and the S 2p core-level spectra are deconvoluted into six peaks with a pseudo-Voigt function by keeping the same FWHM for all the corresponding peaks. The background signals were subtracted using the iterative Tougaard method in CasaXPS software and calibrated using the C 1s peak at 284.6 eV. To fit the Mo 3d spectrum, the separation and the relative area ratio between the two spin-orbit doublets ($3d_{5/2}$ and $3d_{3/2}$) are maintained as 3.1eV and 1.5, respectively. To fit the S 2p spectrum, the separation and the relative area ratio between the two spin-orbit doublets ($2p_{3/2}$ and $2p_{1/2}$) are maintained at 1.2 eV and 2, respectively. As evident, $Mo^{4+}$ valence states predominate in all the samples, which confirms pure $MoS_2$ phase formation. $Mo^{4+}$ $3d_{5/2}$ and $3d_{3/2}$ are located at 229.6 and 232.7 eV, $Mo^{5+}$ $3d_{5/2}$ and $3d_{3/2}$ are located at 230.3 and 233.4 eV, and $Mo^{6+}$ $3d_{5/2}$ and $3d_{3/2}$ are located at 232.8 and 235.9 eV, respectively. Lower valence states of Mo are also present there, which are labeled as 1 and 2, confirming the presence of reduced Mo, i.e., the generation of S-vacancies and the formation of amorphized $MoS_{2-x}$ states at the surface. The lower valence states are situated at 228.6 and 231.7 eV, respectively. As the H0 sample also contains some lower valence states of Mo, it indicates the CVD-grown pristine sample has some defects and S-vacancies. The shoulder peak situated at 226.7 eV corresponds to S-2s[65–69]. With increasing irradiation fluences up to the H2 sample, the peaks corresponding to 1 and 2 are increasing and S-2s are decreasing, indicating the formation of S-vacancies and reduced Mo species at the surface. As $MoS_2$ is a layered structure material with Van der Waals interaction, it tends to absorb oxygen at the surface[21]. Consequently, the pristine sample contains Mo 6+ states. Although Mo 6+ states decreased and Mo 5+ states increased after hydrogen irradiation and annealing, this suggests that in a reductive $H_2$ environment, there is a tendency to convert 6+ to 5+ states (Table S4). Similar behaviour is observed for the H0 to HB samples. It is reported that hydrogen incorporation in the $MoO_3$ matrix and the generation of $H_xMoO_3$ can induce RTFM[33,70]. For the H3 sample, the peak intensity of 1 and 2 decreased and Mo 6+ states increased; the reason is analysed in the next S 2p section elaborately. From the FESEM images, it is evident that the HC samples got damaged and agglomerated at 300°C, the Mo 5+ and 6+ states increased, and the area under peaks 1 and 2 decreased, indicating sputtered S incorporated in the vacancies, causing the diamagnetic behavior of the sample (Figure S8).

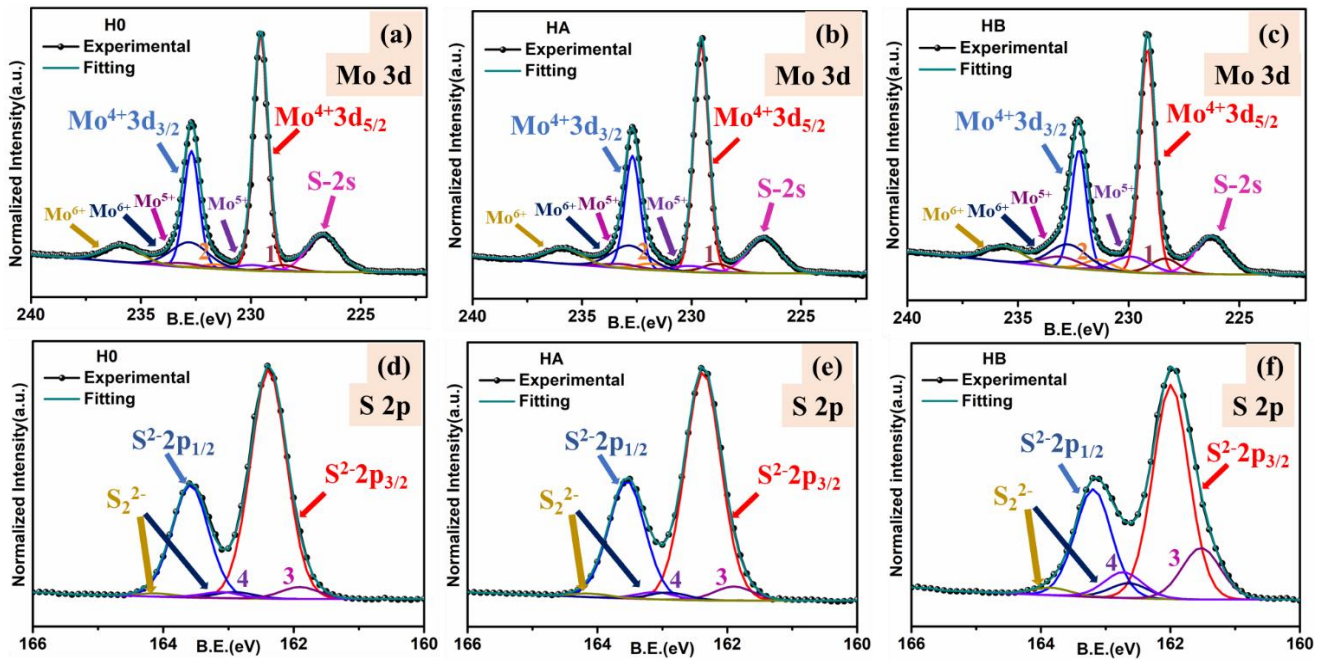

**Figure 8.** Mo 3d core level XPS spectra of (a) Pristine and (b)-(c) $H_2$ annealed samples. The intensities of peaks corresponding to lower valence states increase and the area under Mo 6+ decreases with increasing annealing temperature. S 2p core level spectra of (d) Pristine and (e)-(f) $H_2$ annealed samples. The lower and higher valence states of S is increasing with increasing annealing temperature indicating increasing in polysulfide ions and S-vacancies.

Figures 10 (a) and (b) show the valence band spectra of the pristine, $H_2^+$ irradiated, and $H_2$ annealed samples. The spectra comprise the density of states (DOS), dominated by Mo 4d and S 3p states. The peak indicated by 2 in the Figure predominantly indicates the Mo 4d state ($d_z^2$ state), and the peaks 3,4, and 5 indicate the presence of Mo 4d and S 3p hybridized states[64]. However, it is also reported that peaks 3 and 4 are generated due to the hybridization of O 2p derived states with Mo 4d and S 3p states[71]. From the Mo 3d spectrum, it is already confirmed above that the samples contain some amount of 5+ and 6+ states of Mo. Hence, the peaks 3, 4, and 5 are comprised of Mo 4d, S 3p, and O 2p-derived states. An extra shoulder between 0 and 1 eV is observed for the H2 and HB samples, which corresponds to the lower valence states of Mo and indicates the creation of defects (S-vacancies) and metallic surface[64]. This observation can be correlated to the analysis of the Mo 3d spectrum, wherein it was found that both H2 and HB samples contain higher amounts of lower valence states of Mo.

Moreover, the spectral features corresponding to the occupied Mo 4d-derived states (located at ~2.4 eV) shift towards the Fermi level, resulting in shifting of the valence band maximum (VBM) with increasing irradiation fluences and annealing temperature. These shifts also confirmed band bending or pinning of the Fermi level due to the creation of defects (like peak 3 shifting from 3.6 eV to 3.3 eV). It should also be noticed that at higher irradiation fluences and annealing temperatures, the features corresponding to hybridization of Mo 4d and S 3p-derived states are broadened considerably, indicating that irradiation at higher fluences and annealing at higher temperatures modifies the band structure of $MoS_2$ flake-like films and the generation of amorphous material at the surface[67,72]. Although for H3 and HC samples the VBM shifted away from the Fermi level (inset of Figure 10(a)), this may be attributed to the formation of polysulfide ions and clusters of metallic Mo atoms at such a high fluence, this may result in the filling of vacancies by sulfur and, to a lesser extent, the incorporation of oxygen into the vacancies. All peaks in the VB spectrum of the HC sample are broadened and reduced in intensity, implying significant deterioration of band structure at a temperature of 300 °C.

To investigate the underlying reason for the enhanced magnetism after $H_2$ irradiation, DFT-based theoretical calculations are performed. Bulk $MoS_2$ is a diamagnetic semiconductor, and it has a magnetization of 0.0 $\mu_B$. The creation of an S-vacancy in the bulk $MoS_2$ also shows diamagnetic behavior (Figure S9). In the previous section, it is elaborately explained that in both cases S-vacancies are generated by different mechanisms. From the XPS and EPMA analyses, it is confirmed experimentally that S-vacancies are formed after hydrogen irradiation and annealing. Also, from the FESEM pictures, it is clear that after irradiation and annealing, the number of edge-terminated structures increases and the rough basal plane increases active sites. However, there is also a chance to attach hydrogen with S, enhancing ferromagnetism[61,73]. In this study, calculations are conducted on edge-terminated $MoS_2$ structures, focusing on the effects of S-vacancies and H-adsorption on the magnetization properties. The edge-oriented $MoS_2$ structure, as shown in Figure 11 (a) exhibits a magnetization of 3.2 $\mu_B$, suggesting intrinsic ferromagnetic behavior at room temperature. This magnetization is further enhanced to 3.85 $\mu_B$ when an S-vacancy is introduced at the $MoS_2$ edge (see Figure 11 (b)), aligning well with experimental observations. Additionally, the study investigates the impact of H-adsorption at different positions relative to S atoms. The hydrogen adsorbed initially on the tilt position of S gets optimized to the parallel position as shown in Figure 11 (c) and (d). In this parallel configuration, hydrogen adsorption induced a magnetization of 3.43 $\mu_B$, showing a rise in magnetization compared to the pristine case. Furthermore, we have plotted the partial density of states (PDOS) for $MoS_2$ bulk and edge systems (with and without H adsorption) to give a comparative analysis (Figure S10). In bulk $MoS_2$, the total DOS for up and down spins is symmetrical, indicating no net magnetization and thus diamagnetic behavior. The H adsorption causes the shift of the Fermi level and the appearance of additional states near the Fermi energy. In contrast, the $MoS_2$ edge system is asymmetrical in terms of up and down spins, showing net magnetization in the system. With hydrogen adsorption, there is a shift in the Fermi level, and the PDOS near the Fermi level changes, reflecting modifications in the electronic structure. From these theoretical calculations, it is concluded that edge-oriented $MoS_2$ is itself ferromagnetic at room temperature. Creating S-vacancies and H-adsorption on the S site further enhances the magnetization enriching the material's magnetic properties. The modification of edges (edge-sharpening and roughening) and the creation of S-vacancies occur during both hydrogen irradiation and annealing. The aforementioned discussion also demonstrates that magnetism can be enhanced by H-adsorption with sulfur (S-H) in a parallel position. This scenario is more probable in the context of annealing than in that of irradiation, as hydrogen can be easily incorporated into the matrix due to thermal diffusion. Therefore, the magnetization of $H_2$-annealed samples is significantly enhanced.

**Conclusions:**

Nanostructured $MoS_2$ thin films were grown on a $SiO_2$/Si substrate by the CVD technique. The pristine sample exhibits edge-terminated nanostructures and RTFM with a magnetization value of 1.66 emu/g, suggesting that edge states significantly contribute to the induction of magnetism. Further to generate active sites by introducing defects, low-energy hydrogen ion irradiation, and hydrogen annealing at different temperatures have been performed. For both cases, the magnetization value increases. The highest magnetization value of 2.7 emu/g was obtained for the sample annealed in $H_2$-environment at 200ºC. The XRD and Raman spectroscopy indicate that lattice degradation and the formation of defects occur after post-treatment. FESEM images confirmed the formation of active sites and the sharpening of terminated edges, which further increases the magnetization. EPMA and XPS analyses confirmed the formation of S-vacancies in irradiated and annealed samples up to a particular irradiation dose and annealing temperature. DFT-based theoretical calculation satisfied the experimental results and confirmed the edge-terminated structure, hydrogen adsorption with sulfur, and S-vacancies increase the magnetization value more than the pristine sample. The present study suggests that the presence of edge states along with S-vacancies is important for observing ferromagnetism in nanostructured $MoS_2$ for both cases. As hydrogen adsorption with sulfur in a parallel position enhances the magnetism and is preferable for the annealing cases, H$_2$-annealed samples exhibit a higher magnetization value than the irradiated samples.

**Acknowledgements:**

The authors are grateful to the Physics Department at IIT Delhi for the XRD, Raman, and SQUID facilities; and the Central Research Facilities (CRF) for the FESEM, EPMA, and XPS facilities. The authors also acknowledge the low-energy ion beam facilities provided by IUAC, New Delhi. S. Dey and A. Phutela, acknowledge IIT Delhi for the senior research fellowship and S. Bhattacharya is thankful to the Science and Engineering Research Board (SERB) for the financial support under the core research grant (grant no. CRG/2019/000647).